\documentclass[prd,nofootinbib,preprint,superscriptaddress,twocolumn,10pt]{revtex4}
\pdfoutput=1
\usepackage[T1]{fontenc}
\usepackage{amsmath,amssymb}
\usepackage{braket}
\usepackage{epsfig}
\usepackage{graphicx}
\usepackage[usenames,dvipsnames]{color}
\usepackage{subfigure}
\usepackage{slashed}
\usepackage[colorlinks,citecolor=blue]{hyperref}

\usepackage{pdfpages}
\usepackage{color}
\usepackage{comment}
\usepackage{tikz-feynman}

\begin{document}


\title{ALPy Cogenesis}

\author{Disha Bandyopadhyay}
\email{b.disha@iitg.ac.in}
\affiliation{Department of Physics, Indian Institute of Technology Guwahati, Assam 781039, India}

\author{Debasish Borah}
\email{dborah@iitg.ac.in}
\affiliation{Department of Physics, Indian Institute of Technology Guwahati, Assam 781039, India}
\affiliation{Pittsburgh Particle Physics, Astrophysics, and Cosmology Center, Department of Physics and Astronomy, University of Pittsburgh, Pittsburgh, PA 15260, USA}

\author{Arnab Dasgupta}
\email{arnabdasgupta@pitt.edu}
\affiliation{Pittsburgh Particle Physics, Astrophysics, and Cosmology Center, Department of Physics and Astronomy, University of Pittsburgh, Pittsburgh, PA 15260, USA}

\begin{abstract}
We propose a novel cogenesis scenario by utilising the two-body decay of heavy right-handed neutrino (RHN) via an effective operator involving an axion-like particle (ALP) dark matter (DM) and a light chiral fermion $\nu_R$. This allows the two-body decay of heavy RHN into $\nu_R$ and ALP thereby generating a lepton asymmetry in $\nu_R$ which later gets transferred to left-handed leptons via sizeable Yukawa coupling with a neutrinophilic Higgs doublet. The asymmetry in left-handed leptons is then converted into baryon asymmetry via electroweak sphalerons. The lepton number violation by heavy RHN also induces a one-loop Majorana mass of $\nu_R$ rendering the light neutrinos to be Majorana fermions. Successful leptogenesis constrain the parameter space in terms of RHN mass and axion decay constant. This has interesting consequences for both ALP and QCD axion DM parameter space within reach of several ongoing and near future experiments. We also propose a Dirac version of this scenario without any total lepton number violation. This leads to a long-lived asymmetric Dirac fermion contributing partially to DM thereby opening up more parameter space for ALP. In addition to axion search experiments, the proposed scenarios can have observable signatures at cosmic microwave background (CMB), DM search as well as terrestrial particle physics experiments.
\end{abstract}
\maketitle
\section{Introduction}
The origin of dark matter (DM) and baryon asymmetry in the Universe (BAU) have been longstanding puzzles in particle physics and cosmology \cite{Zyla:2020zbs, Planck:2018vyg}. The failure of the standard model (SM) of particle physics to explain these observed phenomena has led to several beyond standard model (BSM) proposals in the literature. Among them, the weakly interacting massive particle (WIMP) paradigm of DM \cite{Kolb:1990vq, Jungman:1995df} and baryogenesis/leptogenesis \cite{Weinberg:1979bt, Kolb:1979qa, Fukugita:1986hr} are the most popular ones. While the minimal WIMP scenarios are pushed to a tight corner by null results at direct search experiments \cite{LZ:2022lsv}, generic baryogenesis models operate at a high scale out of reach from direct experimental probes. Minimal vanilla leptogenesis scenarios also have a lower bound on the scale of leptogenesis $\geq 10^9$ GeV, known as the Davidson-Ibarra (DI) bound \cite{Davidson:2002qv}. While a low scale realization of leptogenesis with resonantly enhanced CP asymmetry \cite{Pilaftsis:2003gt} exists, some well-motivated DM paradigms beyond the simplest WIMP paradigm have also been explored in the literature. The axion or axion-like particle (ALP) provides such an alternative paradigm for DM, a recent review of which can be found in \cite{OHare:2024nmr}.

The axion \cite{Wilczek:1977pj, Weinberg:1977ma} is the pseudo Nambu-Goldstone boson (pNGB) associated with the Peccei-Quinn (PQ) symmetry $U(1)_{\rm PQ}$ \cite{Peccei:1977hh, Peccei:1977ur, Wilczek:1977pj, Weinberg:1977ma} introduced to solve the strong CP problem, another longstanding puzzle in particle physics. In addition to solving the strong CP problem, an axion with sub-eV mass can constitute the entire cold dark matter in the Universe \cite{Preskill:1982cy, Abbott:1982af, Dine:1982ah}. Axions can also have interesting detection prospects due to their couplings to the SM particles, suppressed by the PQ symmetry breaking scale $f$. For QCD axion models with sub-eV axion mass, there exists a stringent astrophysical bound $f\gtrsim 10^{8}-10^9$ GeV \cite{Raffelt:2006cw, Caputo:2024oqc}. On the other hand, ALP parameter space is less constrained as their couplings and mass can be varied independently. With the growing interest in axion DM and its related phenomenology, it is natural to ask whether this DM paradigm alternative to the WIMP has implications for the observed baryon asymmetry of the Universe.

Motivated by these questions, we propose a novel way of connecting leptogenesis and axion by considering a specific interaction involving axion and suppressed by the PQ scale $f$ to be the origin of CP violation. While there are several ways of connecting the axion with baryogenesis or leptogenesis \cite{Kusenko:2014uta, Servant:2014bla, Adshead:2015jza, Ipek:2018lhm, Croon:2019ugf, Co:2019wyp, Ballesteros:2016xej, Clarke:2015bea, Co:2020jtv, Sopov:2022bog, Domcke:2022kfs, Borah:2024qyo, Mishra:2025twb, Foster:2022ajl}, we for the first time, propose a decay of the type $\psi_1 \rightarrow \psi_2 a$ (with $\psi_{1,2}$ being chiral fermions and $a$ being the axion) to be the origin of the observed CP asymmetry. As a first working example of this mechanism, dubbed as \textit{ALPy cogenesis}, we consider $\psi_1$ to be a heavy right-handed neutrino (RHN) $N_R$ while $\psi_2$ to be a light chiral fermion $\nu_R$, both of which are singlets under the SM gauge symmetry\footnote{In another recent work \cite{Cataldi:2024bcs}, a non-thermal leptogenesis scenario was studied where a heavy ALP decays into RHN in the early Universe. On the other hand, \cite{Datta:2024xhg} considered spontaneous leptogenesis with sub-GeV ALP.}. A non-zero CP asymmetry is generated after taking one-loop correction to this decay if at least two copies of $N_R$ are considered. The asymmetry generated in $\nu_R$ is then transferred to left-handed leptons via sizeable Yukawa interactions with a neutrinophilic Higgs doublet \cite{Heeck:2013vha, Borah:2022qln}. Due to the lepton number violation induced by heavy RHN $N_R$, a Majorana mass of $\nu_R$ is also generated at one-loop level. This radiative correction splits the three Dirac neutrinos at tree level into three active and three sterile Majorana or pseudo-Dirac neutrinos. Keeping the Majorana mass of $\nu_R$ small to control the washout of the asymmetry created from $N_R$ decay also constrains the scale of leptogenesis and other parameters of the model. As a second working example, we consider a setup without lepton number violation which ensures light neutrinos to be of Dirac type. This requires the presence of another Dirac fermion $\chi$ such that the net CP asymmetries in $\nu_R$ and $\chi$ can be zero due to lepton number conservation. Keeping the $\nu_R$ and $\chi$ sectors out of equilibrium allow a net lepton asymmetry to survive till the sphaleron decoupling epoch. The asymmetric $\chi$, if sufficiently long lived, can contribute to DM along with axions. In both the examples, the scale of leptogenesis as well as the PQ scale are constrained from the requirement of generating the observed BAU and DM. The PQ scale, on the other hand, also controls the detection prospects of axion or ALP over a wide range of experiments. The
thermal leptogenesis model explored here can lead to multiple sources
of dark radiation that, taken together, might be within the reach of
current or future cosmic microwave background (CMB) experiments. The
dark radiation can include both axions, which may be thermalized via
the axion portal operator responsible for generating lepton asymmetry, as well as Dirac neutrinos. It should be noted that the presence of ALP or axion in the interaction of heavy neutral fermion is primarily motivated from enhanced detection prospects of the mechanism. For example, leptogenesis can still work if the axion is replaced by a photon, as studied in the context of electromagnetic leptogenesis \cite{Bell:2008fm, Choudhury:2011gbi, Borah:2025oqj} or by a real scalar. 

This paper is organized as follows. In section \ref{sec1}, we discuss the minimal framwork with Majorana neutrinos. In section \ref{sec2}, we discuss the Dirac ALPy cogenesis scenario followed by the detection prospects of both the scenarios in section \ref{sec3}. We finally conclude in section \ref{sec4}.

\section{Minimal Framework}
\label{sec1}
In the minimal realization of the ALPy cogenesis idea, the SM lepton content is extended by two different types of singlet chiral fermions $N_R, \nu_R$. We consider three copies each for these chiral fermions. While $\nu_R$ combines with the left chiral neutrinos $\nu_L$ to form Dirac neutrinos at tree level via a neutrinophilic Higgs doublet $H_2$, the heavy right-handed neutrino $N_R$ has an effective ALP coupling with $\nu_R$. The relevant Lagrangian is given by 
\begin{align}
   -\mathcal{L} & \supset \frac{\partial_\mu a}{f} \overline{N_{R_i}}(\lambda_{i\alpha} \gamma^\mu+\lambda'_{i\alpha} \gamma^\mu \gamma^5) \nu_{R_{\alpha}} + \frac{1}{2} M_i  \overline{N^c_{R_i}} N_{R_i} \nonumber \\
   & + y_{\alpha \beta} \overline{L}_\alpha \tilde{H_2} \nu_{R_\beta} + {\rm h.c.} 
   \label{eq1}
\end{align}
Similar higher dimensional operators with axion interaction being replaced by electromagnetic dipole operator were used for leptogenesis in earlier works \cite{Bell:2008fm, Choudhury:2011gbi, Borah:2025oqj}. While we do not study the details of UV complete models in this work, suitable discrete symmetries can be imposed to realize the above interactions. For example, a $Z_4$ symmetry with the above-mentioned fields charged as $N_{R_i} (-1), \nu_R (i), H_2 (i), a (i)$ can prevent renormalizable Yukawa interactions of $N_{R_i}$ with the SM leptons either via SM-like Higgs $H_1$ or neutrinophilic Higgs $H_2$. Additionally, a soft $Z_4$ breaking term in the scalar potential $\mu^2_{\rm soft} H^\dagger_1 H_2$ leads to an induced vacuum expectation value (VEV) of $H_2$ as
\begin{equation}
    \langle H^0_2 \rangle \equiv v_2\approx\frac{\mu^2_{\rm soft} v_1}{M^2_{H_2}}
\end{equation}
where $v_1 $ is the VEV of the SM-like Higgs $H_1$ and $M_{H_2}$ is the bare mass of $H_2$. This also leads to mixing between neutral scalar of $H_2$ and the SM Higgs resulting in interesting phenomenology which has been studied in several earlier works \cite{Maitra:2014qea, Haba:2011nb, Gabriel:2008es, Seto:2015rma, Bertuzzo:2015ada, Huitu:2017vye}.

\begin{figure}
    \includegraphics[scale=0.4]{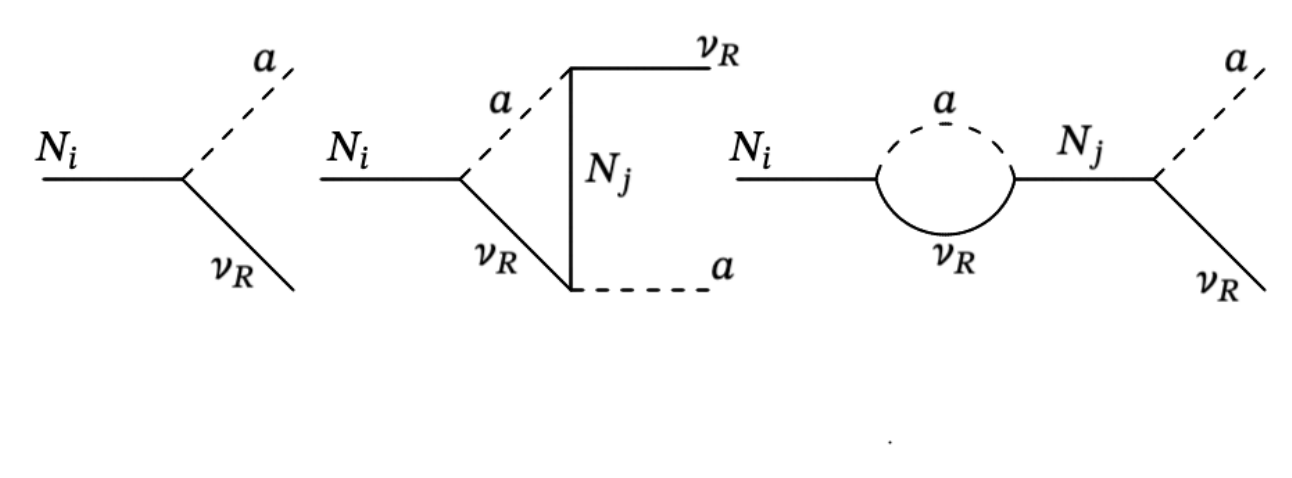}
    \caption{Processes responsible for creating asymmetry in $\nu_R$.}
    \label{fig1}
\end{figure}

Fig. \ref{fig1} shows the two-body decay $N_i \rightarrow \nu_{R_\alpha} a$ and possible one-loop corrections responsible for generating the CP asymmetry. The corresponding CP asymmetry, where the lepton flavor $\alpha$ of $\nu_{R_\alpha}$ is being summed over, can be written as
\begin{align}
    \epsilon_i & = \frac{\Gamma (N_i \rightarrow \nu_R a) - \Gamma (N_i \rightarrow \overline{\nu_R} a)}{\Gamma (N_i \rightarrow \nu_R a) + \Gamma (N_i \rightarrow \overline{\nu_R} a)} \nonumber \\
    & =\frac{\Gamma (N_i \rightarrow \nu_R a) - \Gamma (N_i \rightarrow \overline{\nu_R} a)}{\Gamma_{i}},
\end{align}
with the tree level decay width being given as
\begin{widetext}
\begin{align}
    \Gamma_0=\Gamma (N_i \rightarrow \nu_{R\alpha} a) = \Gamma (N_i \rightarrow \overline{\nu_{R\alpha}} a) &= \frac{|\widetilde{\lambda}_{i\alpha}|^2 M_i}{32\pi}\left(\frac{M_i}{f}\right)^2 \left(1+r^4_a-r^2_a(1+r^2_\alpha) - 2r^2_\alpha \right)\Lambda^{1/2}(1,r^2_a,r^2_\alpha).
    \end{align}
\end{widetext}
In the above expression
\begin{equation}
    \Lambda(x,y,z) = x^2 + y^2 + z^2 -2xy - 2xz - 2yz, \nonumber \\
    \end{equation}
and $r_a = m_a/M_i, r_{\alpha,\beta} = m_{\nu_{\alpha,\beta}}/M_i$, $\widetilde{\lambda}_{i \alpha} = \lambda_{i \alpha} + \lambda'_{i\alpha}$. We always work in the limit where $r_a, r_{\alpha} \ll 1$ such that $\Gamma_0 \approx \frac{|\widetilde{\lambda}_{i\alpha}|^2 M_i}{32\pi}\left(\frac{M_i}{f}\right)^2$. The CP asymmetry parameter, considering self-energy correction, is given by
\begin{align}
    \epsilon_i = \frac{{\rm Im}[\widetilde{\lambda}_{j\alpha}\widetilde{\lambda}_{j\beta}\widetilde{\lambda}^*_{i\beta}\widetilde{\lambda}^*_{i\alpha}]}{|\widetilde{\lambda}_{i\beta}|^2|\widetilde{\lambda}_{i\alpha}|^2 M_i}\Gamma_0(N_i \rightarrow \nu_{R\beta} a)\frac{r_{ij}}{1-r^2_{ij}}
\end{align}
where $r_{ij} = M_j/M_i$. Now for the resonant regime $\Delta M = M_i - M_j = \Gamma_0/2$ then the CP asymmetry becomes maximal $\epsilon \sim \sin{(2\phi)}$ with $\phi$ being the relative phase. In order to arrive at this expression, $\lvert \tilde{\lambda} \rvert$ is assumed to be the common magnitude of the couplings $\tilde{\lambda}_{i\alpha}=\lvert \tilde{\lambda} \rvert e^{i\theta_{i\alpha}}$ with a common relative phase angle given by $\phi=\theta_{i\alpha}-\theta_{j \alpha}$.

The asymmetry generated in $\nu_R$ is then transferred to the left-handed leptons via sizeable coupling with $H_2$. If this asymmetry is transferred to the left-handed leptons before sphalerons decouple, a non-zero baryon asymmetry can be generated. The neutrinos acquire small Dirac mass at tree level from the VEV of the neutral component of $H_2$ as
\begin{equation}
    M_D = \frac{y v_2}{\sqrt{2}}
\end{equation}
where $v_2$ can be much smaller than the VEV of the SM-like Higgs doublet $H_1$. This can be achieved if the VEV of $H_2$ is protected by an approximate global symmetry like the $Z_4$ symmetry mentioned above and can only be induced after electroweak symmetry breaking due to the VEV of $H_1$. Smallness of $v_2$ compared to $v_1 \equiv \langle H_1 \rangle$ ensures that the Yukawa coupling $y$ is sufficiently large to transfer the asymmetry from $\nu_R$ to $\nu_L$ as we discuss below. 

\begin{figure}[h]
    \includegraphics[scale=0.5]{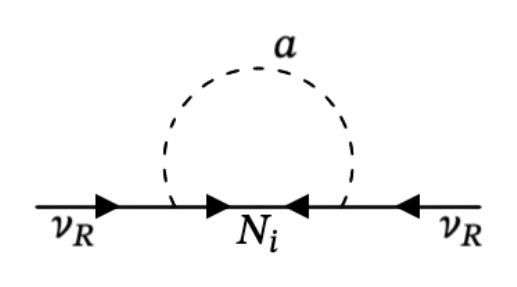}
    \caption{One-loop contribution to Majorana mass of $\nu_R$.}
    \label{fig2}
\end{figure}

One can also have a Majorana mass term of $\nu_R$ generated at one loop as shown in Fig. \ref{fig2}. The Majorana mass of $\nu_R$ can be estimated as
\begin{equation}
    M^{\nu_R}_{\alpha \beta} \approx \frac{1}{16\pi^2} \frac{1}{f^2} \tilde{\lambda}_{i \alpha} M^3_{i} \tilde{\lambda}_{i \beta}. 
\end{equation}
A large Majorana mass of $\nu_R$ can introduce additional source of lepton asymmetry and washout of the asymmetry already produced from $N_i$ decay. As this Majorana mass term is very similar to the decay width of $N_i$, keeping it small also forces $\Gamma_{i}$ to be small, which leads to interesting correlations among the scale of leptogenesis and light sterile neutrino masses. Note that this similarity between decay width of $N_i$ and Majorana mass of $\nu_R$ holds only in the absence of any specific texture of $\tilde{\lambda}$ or strong phase-space suppression in $\Gamma_0$. In our analysis, we are neither using any specific textures nor phase-space suppression and we explicitly calculate them without assuming any similarities.

\subsection{Leptogenesis}

While there are three heavy neutral fermions $N_i$, the lepton asymmetry is generated by out-of-equilibrium decay of the lightest among them. The relevant Boltzmann equations can be written as 
\begin{equation}
  \dfrac{d \eta_{\rm N_{1}}}{d z}= -D_{1} \left( \eta_{\rm N_{1}}-\eta_{\rm N_{1}}^{\rm eq} \right),
  \label{eq:BE22}
  \end{equation}
\begin{equation}
 \dfrac{d \eta_{\rm B-L}}{d z}= -\epsilon_{1} D_{1} (\eta_{\rm N_{1}}-\eta_{\rm N_{1}}^{\rm eq})-W \,\eta_{\rm B-L},
\end{equation}
where $\eta_x=n_x/n_\gamma$ is the comoving number density of $x$ and $z=M_1/T, D_1= \langle \Gamma_{1} \rangle/(\mathcal{H}z)$ with $\mathcal{H}$ being the Hubble expansion parameter. The thermal averaged decay width is given by $\langle \Gamma_{1} \rangle = \Gamma_{1} K_1 (z)/K_2 (z)$ where $K_i$'s denote modified Bessel functions of the second kind. $W$ denotes the rates of washout processes including inverse decay and scatterings. The inverse decay rate can be written in terms of the decay rate as $\Gamma_{\rm ID}= \Gamma_{1} (z) \frac{\eta_{\rm N_1}^{\rm eq} (z)}{\eta_{\rm l}^{\rm eq}(z)}$, which gives the washout term due to inverse decay $W_{\rm ID}$ as
\begin{align}
    W_{\rm ID}(z)=\frac{1}{4} K z^3 K_1 (z).
\end{align}
$K$ denotes the decay parameter defined as
\begin{align}
    K=\frac{\Gamma_{1}}{\mathcal{H} (z=1)}.
\end{align}
The washout term due to inverse decay can be approximately written as \cite{Buchmuller:2004nz}
\begin{align}
    W_{\rm ID}(z)\simeq\frac{1}{4} K z^2 \sqrt{1+\frac{\pi}{2}z}e^{-z}\,\label{eqn:wid2}
\end{align}
which is in equilibrium when $W_{\rm ID}(z)\geq1$.

The scattering processes leading to washout include $\nu_R \nu_R \leftrightarrow a a$ as one of the dominant ones. The requirement for keeping this washout process (mediated by $N_1$) out of equilibrium leads to
\begin{equation}
   \Gamma^{\rm wo}_1 \equiv n^{\rm eq} \sigma v < \mathcal{H} \implies T \lesssim \frac{f^4}{M_{\rm Pl} M^2_1 (\tilde{\lambda}^\dagger \tilde{\lambda})^2_{11}}. 
\end{equation}
Adopting a conservative approach of demanding this washout process to be out of equilibrium at $T \leq M_1$ leads to
\begin{equation}
 f \gtrsim \sqrt{\Gamma_{1} M_{\rm Pl}}.
\end{equation}
For $f > M_1$, the above condition also ensures $K<1$, keeping the washout under control. Another important washout process is $L L \leftrightarrow H_2 H_2$ or other variants like $L H^\dagger_2 \leftrightarrow \overline{L} H_2$ which violate lepton number by two units. Demanding this washout to be out of equilibrium $(\Gamma^{\rm wo}_2 < \mathcal{H})$ leads to
\begin{equation}
    \frac{y^4 (M^{\nu_R})^2}{T^4} T^3 \lesssim \frac{T^2}{M_{\rm Pl}}.
\end{equation}
Using a conservative approach to ensure this washout to be out-of-equilibrium for $T \gtrsim T_{\rm sph} \approx 130$ GeV leads to an upper bound on the Dirac Yukawa coupling as 
\begin{equation}
    y < \frac{T^{3/4}_{\rm sph}}{M^{1/4}_{\rm Pl} \Gamma^{1/2}_{1}}
\end{equation}
which, for $\Gamma_{1} \sim 1$ keV, leads to $y < \mathcal{O}(1)$. For larger decay width of $N_1$ and hence larger Majorana mass of $\nu_R$, the upper bound will be stronger. The Dirac Yukawa coupling can not be arbitrarily small as the asymmetry in $\nu_R$ gets transferred to the left sector via this Yukawa portal interactions only. Demanding this Yukawa interaction to be in equilibrium prior to the sphaleron decoupling epoch leads to a lower bound 
\begin{equation}
    y \gtrsim 10^{-8}
\end{equation}
which gets stronger if the Yukawa interactions are required to be in equilibrium at higher temperatures. However, as far as the generation of baryon asymmetry is concerned, it is sufficient to ensure that this process enters equilibrium before the sphalerons decouple. We are also not assuming any specific flavor structure of the Dirac Yukawa coupling $y$ in our analysis, assuming it to be flavor universal. A full treatment of flavored leptogenesis \cite{Abada:2006fw,Nardi:2006fx,Abada:2006ea,Blanchet:2006be, BhupalDev:2014pfm, Dev:2017trv} is beyond the scope of the present work.

Once the asymmetry in $\nu_R$ is transferred to the left-handed lepton doublets, the electroweak sphaleron processes convert the $B-L$ asymmetry into baryon asymmetry as
\begin{align}
   \eta_{\rm B}= \frac{8 N_f + 4 N_\textbf{H}}{22 N_f + 13 N_\textbf{H}}\frac{\eta_{\rm B-L}}{S} = \frac{C_{\rm sph}}{S} \eta_{\rm B-L}\,,\label{eqn:sphconv}   
\end{align}
where $C_{\rm sph}= \frac{8}{23}$ for two Higgs doublets $N_\textbf{H}=2$ and three fermion generations $N_f=3$. The factor $S$ accounts for the change in the relativistic degrees of freedom from the scale of leptogenesis until recombination and comes out to be $S=\frac{106.75}{3.91}\simeq27.3$. The final baryon asymmetry $\eta_{\rm B}$ can also be estimated analytically in terms of the CP asymmetry parameter as \cite{Buchmuller:2004nz}
 \begin{align}
     \eta_B = \frac{C_{\rm sph}}{S} \epsilon_1 \kappa\,, \label{eqn:etaana}
 \end{align}
where $\kappa$ is known as the efficiency factor which incorporates the effects of washout processes. Instead of considering any approximate value of $\kappa$ valid for a specific washout regime, we solve the Boltzmann equations numerically to calculate the asymmetry explicitly.

\begin{figure}
    \includegraphics[scale=0.5]{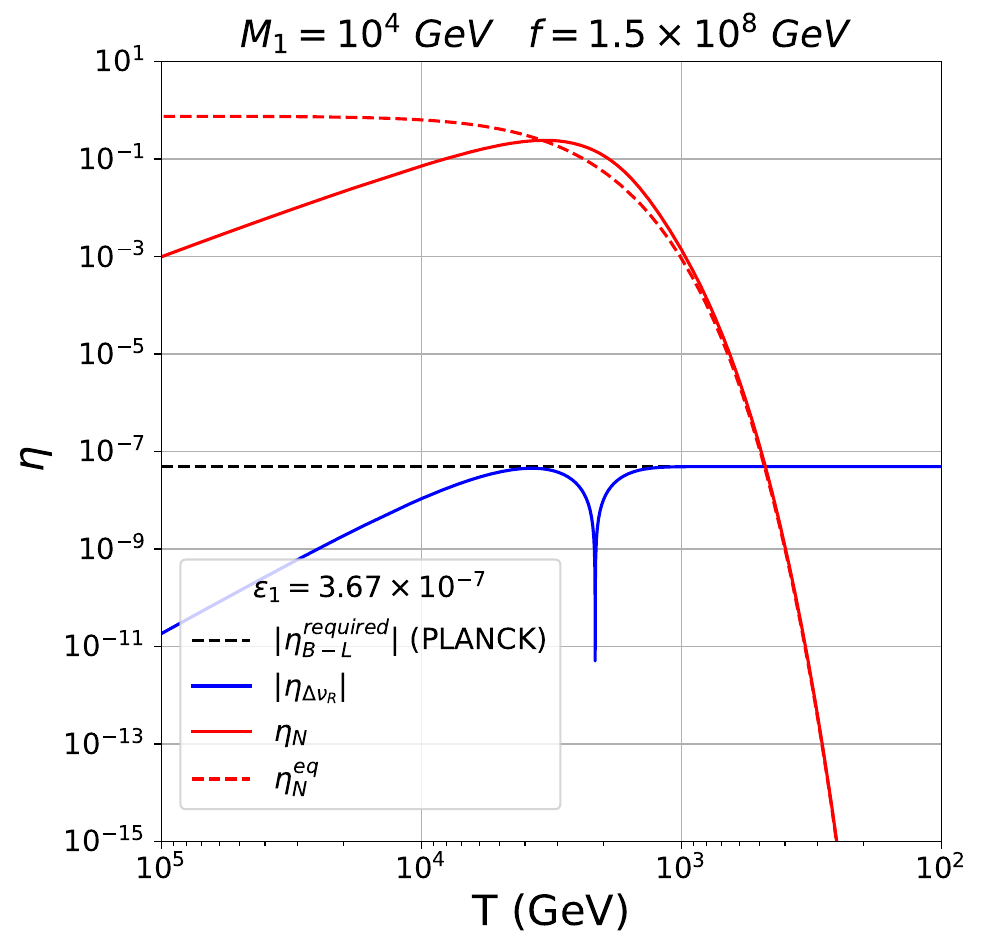}
    \caption{Evolution of comoving number densities in minimal ALPy cogenesis.}
    \label{fig2a}
\end{figure}

\begin{figure}
    \includegraphics[scale=0.3]{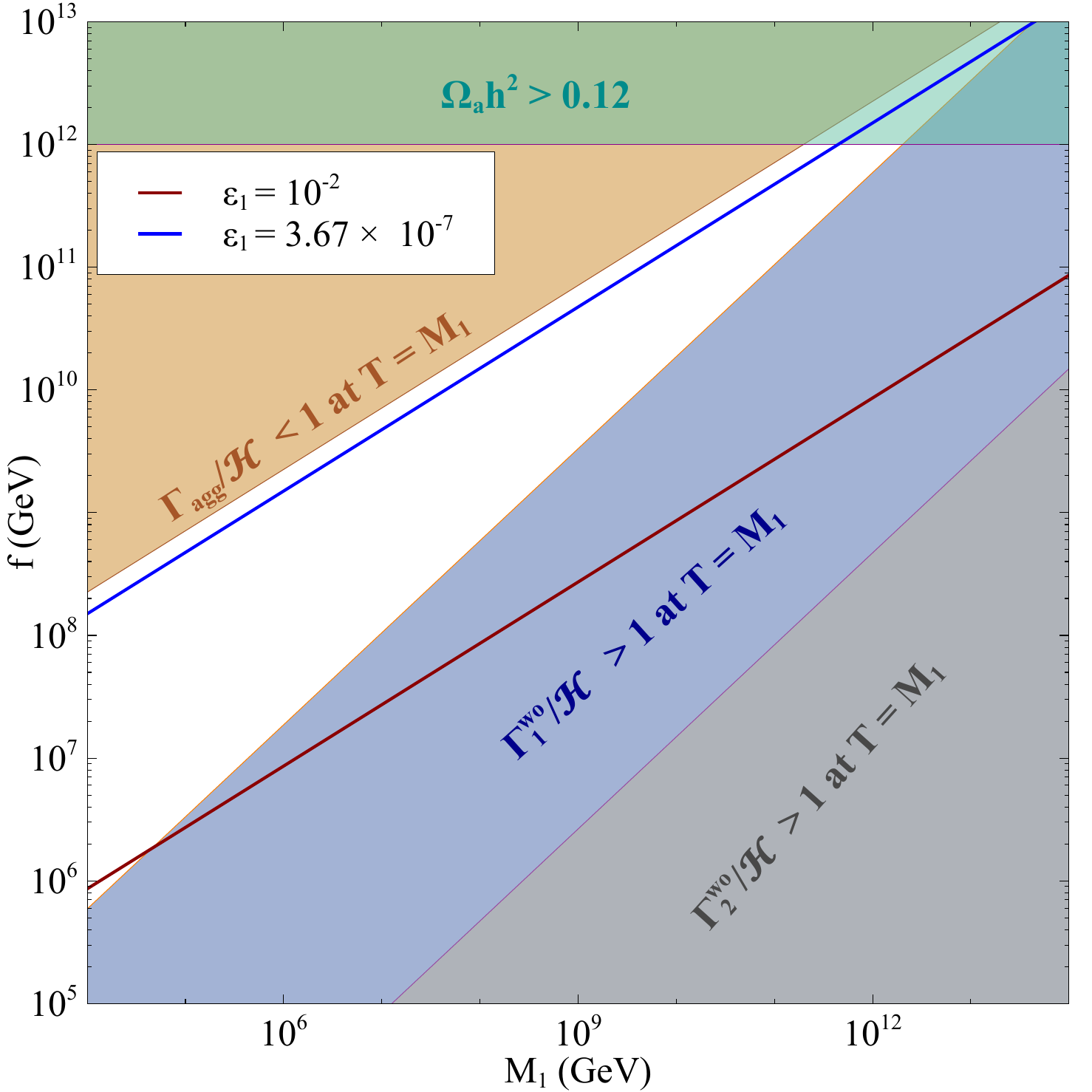}
    \caption{Allowed parameter space of minimal ALPy cogenesis scenario in $f-M_1$ plane considering dimensionless couplings $\tilde{\lambda}_{i\alpha}, y_{\alpha \beta}\sim 10^{-2}$. The shaded regions are disfavored leaving only the white region allowed for minimal ALPy cogenesis.}
    \label{fig2b}
\end{figure}
Fig. \ref{fig2a} shows the evolution of comoving number densities of heavy RHN, and the absolute value of $\nu_R$ asymmetry. As the RHN abundance gradually rises, the lepton asymmetry gets a sign-flip leading to the resonance type feature. The lepton asymmetry saturates subsequently to the required value and transferred into asymmetry in lepton doublets which are in equilibrium with $\nu_R$ via neutrinophilic Higgs doublet $H_2$. The asymmetry in lepton doublet is then converted into baryon asymmetry when electroweak sphalerons go out of equilibrium at $T=T_{\rm sph}$. Fig. \ref{fig2b} shows the parameter space consistent with successful minimal ALPy cogenesis in $f-M_1$ plane for two different values of the CP asymmetry parameter $\epsilon_1$. The points on the blue and maroon colored solid lines are consistent with the observed baryon asymmetry for two different values of the CP asymmetry parameter $\epsilon_1 = 3.67 \times 10^{-7}, \epsilon_1 = 10^{-2}$ respectively. The upper left triangular part of the plane labeled as $\Gamma_{\rm agg}/\mathcal{H} <1$ corresponds to the region where the axion-gluon interactions are not in thermal equilibrium \cite{Masso:2002np}. Since we are assuming $a, \nu_R$ to be part of the bath while solving the Boltzmann equations, our analysis remains valid outside this shaded region only. The calculation of the baryon asymmetry with non-thermal axions require solving the Boltzmann equation for axions and estimating the corresponding non-thermal abundance of heavy neutral fermions. We do not pursue such an analysis here and stick to the thermal leptogenesis regime only. The shaded regions labeled as $\Gamma^{\rm wo}_{1,2}/\mathcal{H} > 1$ are disfavored due to strong washout rates. The horizontal shaded region at the top corresponds to the parameter space where the QCD axion DM is overproduced via the vacuum misalignment mechanism \cite{Preskill:1982cy, Abbott:1982af, Dine:1982ah}. It should be noted that the dimensionless couplings $\tilde{\lambda}_{i\alpha}, y_{\alpha \beta}$ are assumed to be $\sim 10^{-2}$ in Fig. \ref{fig2b}. The constraints from washouts become stronger for larger values of these dimensionless couplings, shrinking the allowed white colored region in Fig. \ref{fig2b}.

\section{Dirac ALPy cogenesis}
\label{sec2}
We now propose a lepton number conserving version of the ALPy cogenesis discussed above. We consider a scenario where the standard model fermion content is extended by two types of vector-like singlet neutral fermions $N_{L,R}, \chi_{L,R}$ and a singlet chiral fermion $\nu_R$. While $\nu_R$ combines with the left chiral neutrinos $\nu_L$ to form Dirac neutrinos at tree level via a neutrinophilic Higgs doublet $H_2$, the heavy vector-like fermion $N$ has an effective ALP coupling with $\nu_R$ as well as $\chi$. A global $U(1)_D$ symmetry, similar to lepton number, is assumed to be conserved which prevents any Majorana terms of singlet fermions. Given this, the relevant Lagrangian invariant under the SM gauge symmetry and $U(1)_D$ is given by 

\begin{align}
   -\mathcal{L} & \supset \frac{\partial_\mu a}{f} \overline{N_{R_i}}(\lambda_{i\alpha} \gamma^\mu+\lambda'_{i\alpha} \gamma^\mu \gamma^5) \nu_{R_{\alpha}} \nonumber \\
  & +  \frac{\partial_\mu a}{f} \overline{N_{R_i}}(h_{Ri} \gamma^\mu+h'_{Ri} \gamma^\mu \gamma^5) \chi_{R} \nonumber \\
  & +\frac{\partial_\mu a}{f} \overline{N_{L_i}}(h_{Li} \gamma^\mu+h'_{Li} \gamma^\mu \gamma^5) \chi_{L} \nonumber \\
   & +y_{\alpha \beta} \overline{L}_\alpha \tilde{H_2} \nu_{R_\beta}+ M_i \overline{N_{iL}} N_{iR} + m_\chi \overline{\chi_L} \chi_R+ {\rm h.c.} 
   \label{eq2}
\end{align}

\begin{figure}
    \includegraphics[scale=0.5]{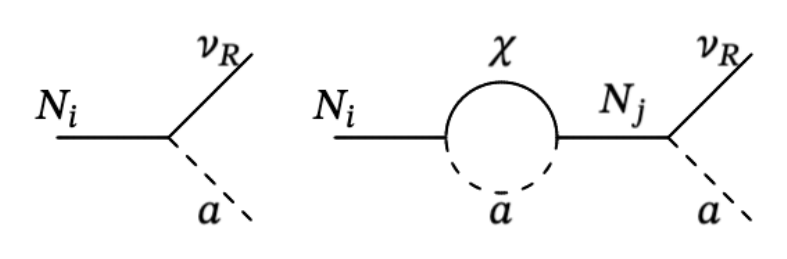}
    \caption{Processes responsible for creating asymmetry in $\nu_R$ in Dirac ALPy cogenesis.}
    \label{fig3}
\end{figure}

The heavy vector-like fermion can decay out-of-equilibrium to create asymmetries in $\nu_R, \chi_L, \chi_R$. This is similar to the idea of Dirac leptogenesis \cite{Dick:1999je, Murayama:2002je, Cerdeno:2006ha} where equal and opposite CP asymmetries are created in left and right-handed neutrino sectors. The processes responsible for creating asymmetry in $\nu_R$ are shown in Fig. \ref{fig3}. Denoting the CP asymmetries as $\epsilon_{\nu_R}, \epsilon_{\chi_L}, \epsilon_{\chi_R}$ and given the fact that there is no net lepton number violation due to pure Dirac nature of all the fermions, we have 
\begin{equation}
    \epsilon_{\nu_R}+\epsilon_{\chi_L}+\epsilon_{\chi_R}=0.
\end{equation}
The CP asymmetry $\epsilon_x$ is defined as
\begin{align}
    \epsilon_x = \frac{\Gamma (N_i \rightarrow x a) - \Gamma (N_i \rightarrow \overline{x} a)}{\Gamma_{i}}.
\end{align}
The CP asymmetries, considering one loop self-energy corrections, are found to be 
\begin{align}
    \epsilon_{\nu_R} &= \frac{{\rm Im}[\widetilde{\lambda}^*_{j\alpha}\widetilde{h}_{Lj}\widetilde{h}^*_{Li}\widetilde{\lambda}_{i\alpha}]}{|\widetilde{h}_{Li}|^2\widetilde{\Gamma}_{\rm tot} M_i}\Gamma_0(N_i \rightarrow \chi_L a)\frac{r_{ij}}{1-r^2_{ij}} \nonumber \\
    &+ \frac{{\rm Im}[\widetilde{\lambda}^*_{j\alpha}\widetilde{h}_{Rj}\widetilde{h}^*_{Ri}\widetilde{\lambda}_{i\alpha}]}{|\widetilde{h}_{Ri}|^2\widetilde{\Gamma}_{\rm tot} M_i}\Gamma_0(N_i \rightarrow \chi_R a)\frac{1}{1-r^2_{ij}},
\end{align}
\begin{align}
    \epsilon_{\chi_R} &= \frac{{\rm Im}[\widetilde{h}^*_{Rj} \widetilde{h}_{Lj} \widetilde{h}^*_{Li}\widetilde{h}_{Ri}]}{|\widetilde{h}_{Li}|^2\widetilde{\Gamma}_{\rm tot} M_i}\Gamma_0(N_i \rightarrow \chi_L a)\frac{r_{ij}}{1-r^2_{ij}} \nonumber \\
    &+ \frac{{\rm Im}[\widetilde{h}^*_{Rj} \widetilde{\lambda}_{j\alpha} \widetilde{\lambda}^*_{i\alpha}\widetilde{h}_{Ri}]}{|\widetilde{\lambda}_{i\alpha}|^2 \widetilde{\Gamma}_{\rm tot} M_i}\Gamma_0(N_i \rightarrow \nu_R a)\frac{1}{1-r^2_{ij}},
\end{align}
\begin{align}
    \epsilon_{\chi_L} &= \frac{{\rm Im}[\widetilde{h}^*_{Lj}\widetilde{h}_{Rj}\widetilde{h}^*_{Ri}  \widetilde{h}_{Li}]}{|\widetilde{h}_{Ri}|^2\widetilde{\Gamma}_{\rm tot} M_i}\Gamma_0(N_i \rightarrow \chi_R a)\frac{r_{ij}}{1-r^2_{ij}} \nonumber \\
    &+ \frac{{\rm Im}[\widetilde{h}^*_{Lj}\widetilde{\lambda}_{j\alpha}\widetilde{\lambda}^*_{i\alpha}\widetilde{h}_{Li}]}{|\widetilde{\lambda}_{i\alpha}|^2\widetilde{\Gamma}_{\rm tot} M_i}\Gamma_0(N_i \rightarrow \nu_R a)\frac{r_{ij}}{1-r^2_{ij}},
\end{align}
where $\widetilde{h}_{Li} = h_{Li}+h'_{Li}$ and $\widetilde{\Gamma}_{\rm tot} = |\widetilde{h}_{Ri}|^2 + |\widetilde{h}_{Li}|^2 + |\widetilde{\lambda}_{i\alpha}|^2$ with other definitions being same as before. Using the above expressions for CP asymmetries, it is clear that $\epsilon_{\nu_R}+\epsilon_{\chi_L}+\epsilon_{\chi_R}=0$ from the requirement of total lepton number conservation.

The Boltzmann equations can be written as 
\begin{align}
   \dfrac{d \eta_{\Delta \nu_R}}{d z} &= \frac{1}{n_\gamma \mathcal{H}z} \bigg [ \epsilon_{\nu_R} \left ( \frac{\eta_{N_{1R}}}{\eta^{\rm eq}_{N_{1R}}}-1 \right ) \gamma (N_{1R} \rightarrow a \nu_R) \nonumber \\
   & -\frac{\eta_{\Delta \nu_R}}{2\eta^{\rm eq}_{\nu_R}} \gamma (N_{1R} \rightarrow a \nu_R) \nonumber \\
   & + \left ( \frac{\eta_{\Delta \chi_R}}{\eta^{\rm eq}_{\chi_R}}-\frac{\eta_{\Delta \nu_R}}{\eta^{\rm eq}_{\nu_R}} \right) \gamma (\chi_R a \rightarrow \nu_R a) \nonumber \\
   & + \left ( \frac{\eta_{\Delta \chi_L}}{\eta^{\rm eq}_{\chi_L}}-\frac{\eta_{\Delta \nu_R}}{\eta^{\rm eq}_{\nu_R}} \right) \gamma (\chi_L a \rightarrow \nu_R a) \bigg ],
\end{align}
\begin{align}
   \dfrac{d \eta_{\Delta \chi_R}}{d z} &= \frac{1}{n_\gamma \mathcal{H}z} \bigg [ \epsilon_{\chi_R} \left ( \frac{\eta_{N_{1R}}}{\eta^{\rm eq}_{N_{1R}}}-1 \right ) \gamma (N_{1R} \rightarrow a \chi_R) \nonumber \\
   & -\frac{\eta_{\Delta \chi_R}}{2\eta^{\rm eq}_{\chi_R}} \gamma (N_{1R} \rightarrow a \chi_R) \nonumber \\
   & + \left ( \frac{\eta_{\Delta \nu_R}}{\eta^{\rm eq}_{\nu_R}}-\frac{\eta_{\Delta \chi_R}}{\eta^{\rm eq}_{\chi_R}} \right) \gamma (\nu_R a \rightarrow \chi_R a) \nonumber \\
   & + \left ( \frac{\eta_{\Delta \chi_L}}{\eta^{\rm eq}_{\chi_L}}-\frac{\eta_{\Delta \chi_R}}{\eta^{\rm eq}_{\chi_R}} \right) \gamma (\chi_L a \rightarrow \chi_R a) \bigg ],
\end{align}
\begin{align}
   \dfrac{d \eta_{\Delta \chi_L}}{d z} &= \frac{1}{n_\gamma \mathcal{H}z} \bigg [ \epsilon_{\chi_L} \left ( \frac{\eta_{N_{1L}}}{\eta^{\rm eq}_{N_{1L}}}-1 \right ) \gamma (N_{1L} \rightarrow a \chi_L) \nonumber \\
   & -\frac{\eta_{\Delta \chi_L}}{2\eta^{\rm eq}_{\chi_L}} \gamma (N_{1L} \rightarrow a \chi_L) \nonumber \\
   & + \left ( \frac{\eta_{\Delta \nu_R}}{\eta^{\rm eq}_{\nu_R}}-\frac{\eta_{\Delta \chi_L}}{\eta^{\rm eq}_{\chi_L}} \right) \gamma (\nu_R a \rightarrow \chi_L a) \nonumber \\
   & + \left ( \frac{\eta_{\Delta \chi_R}}{\eta^{\rm eq}_{\chi_R}}-\frac{\eta_{\Delta \chi_L}}{\eta^{\rm eq}_{\chi_L}} \right) \gamma (\chi_R a \rightarrow \chi_L a) \bigg ],
\end{align}
\begin{align}
    \dfrac{d \eta_{N_{1L}}}{d z} &= -\frac{1}{n_\gamma \mathcal{H}z} \bigg [ \left ( \frac{\eta_{N_{1L}}}{\eta^{\rm eq}_{N_{1L}}} -1 \right) \gamma (N_{1L} \rightarrow a \chi_L) \bigg ],
\end{align}
\begin{align}
    \dfrac{d \eta_{N_{1R}}}{d z} &= -\frac{1}{n_\gamma \mathcal{H}z} \bigg [ \left ( \frac{\eta_{N_{1R}}}{\eta^{\rm eq}_{N_{1R}}} -1 \right) \gamma (N_{1R} \rightarrow a \chi_R) \nonumber \\
    & + \left ( \frac{\eta_{N_{1R}}}{\eta^{\rm eq}_{N_{1R}}} -1 \right) \gamma (N_{1R} \rightarrow a \nu_R) \bigg ],
\end{align}
where $\eta_x=n_x/n_\gamma$, $\eta_{\Delta x}=(n_x-n_{\bar{x}})_/n_\gamma$ and $z=M_1/T$.
In the above equations, the thermally-averaged reaction densities for the $2-2$ scattering processes are defined as 
\begin{eqnarray}
    \gamma_{ij\longrightarrow kl } = \frac{ T}{64 \pi^{4}} \int_{(m_{i}+m_{j})^{2}}^{\infty} ds \sqrt{s}K_{1} \left( \sqrt{s}/T \right) \hat{\sigma}(s),
\end{eqnarray}
where $\hat{\sigma}(s)$ is the reduced cross-section for the process and is given by 
\begin{equation}
\hat{\sigma} (s)= 8 \left[ (p_{i}.p_{j})^{2}-m_{i}^{2}m_{j}^{2} \right] \sigma (s).
\end{equation}
The reaction rate density $\gamma_{ij \longrightarrow kl}$ is related to $\langle  \sigma v \rangle_{ij\longrightarrow kl}$ as 
\begin{equation}
\gamma_{ij \longleftarrow kl}=n_{i}^{\rm eq}n_{j}^{\rm eq}\langle  \sigma v \rangle_{ij \longrightarrow kl}
\end{equation}
where $n_{i}^{\rm eq}$ and $n_{j}^{\rm eq}$ are the equilibrium number densities of species $i$ and $j$ respectively.  For the decay $A\longrightarrow B$, the reaction density is defined as 
\begin{equation}
    \gamma_{A\longrightarrow B}=n_{A}^{\rm eq} \frac{K_{1}\left( m_{A}/T \right)}{K_{2} \left(  m_{B}/T \right)}\Gamma_{A\longrightarrow B}.
\end{equation}
In order to prevent the asymmetry in $\nu_R$ from being washed out, one has to ensure that the process $\nu_R a \leftrightarrow \chi a$ does not equilibrate. This can be taken care of by keeping the rates of this process below the Hubble expansion rate of the Universe $\mathcal{H}$. This leads to the following constraint
\begin{equation}
    \frac{\lvert \tilde{\lambda}_{1 \alpha} \rvert^2 \lvert \tilde{h}_{L 1} \rvert^2  M^2_1}{f^4} T^3 \lesssim \frac{T^2}{M_{\rm Pl}}.
\end{equation}
Due to the lepton number conservation, neither $\nu_R$ acquires any Majorana mass nor there exist any lepton number violating washout processes in addition to inverse decay and left-right equilibration mentioned above.

\begin{figure}
    \includegraphics[scale=0.5]{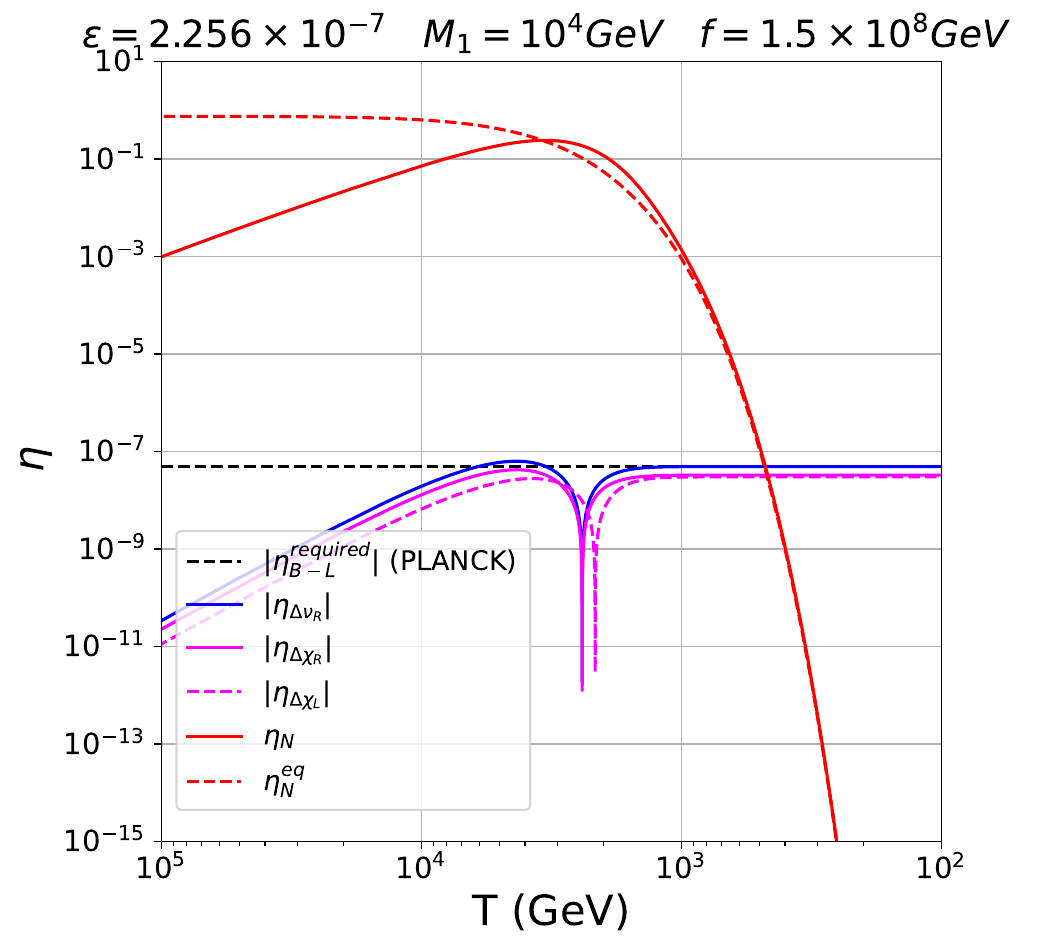}
    \caption{Evolution of comoving number densities in Dirac ALPy cogenesis.}
    \label{fig4}
\end{figure}

\begin{figure}
    \includegraphics[scale=0.3]{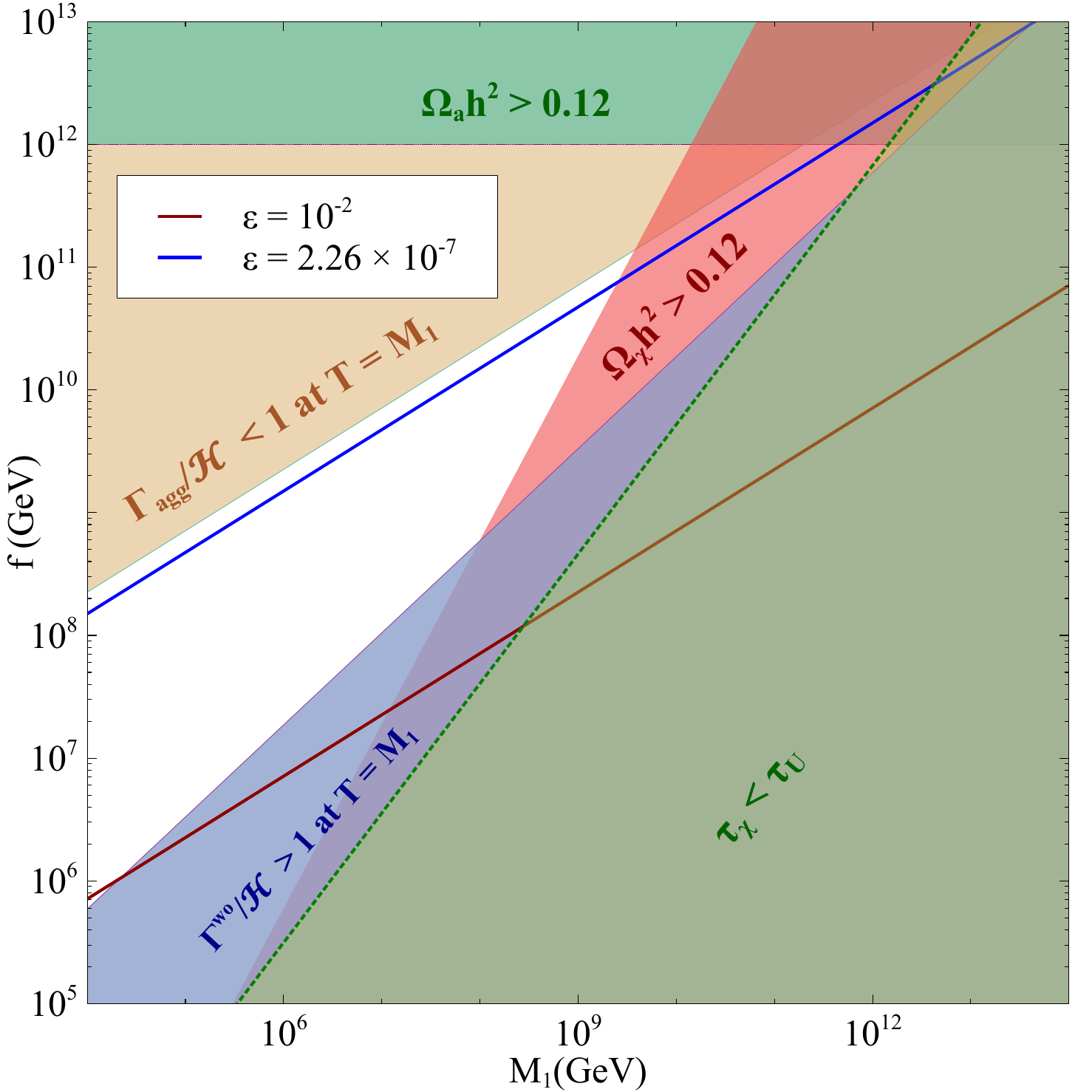}
    \caption{Allowed parameter space of Dirac ALPy cogenesis scenario in $f-M_1$ plane considering Dimensionless couplings $\tilde{\lambda}_{i\alpha}, \tilde{h}_{Li}, \tilde{h}_{Ri}\sim 10^{-2}$. The shaded regions are disfavored due to different constraints, leaving only the white region allowed for Dirac ALPy cogenesis.}
    \label{fig5}
\end{figure}


This leads to the generation of asymmetries in $\nu_R$ and $\chi$ as $N_i$ decays out-of-equilibrium. Similar to the minimal scenario discussed before, the asymmetry in $\nu_R$ gets transferred to the left-handed lepton doublets via Yukawa couplings with the neutrinophilic Higgs doublet. The lepton asymmetry is then converted into baryon asymmetry via the electroweak sphalerons. The asymmetry stored in $\chi$ can survive as asymmetric dark matter. Fig. \ref{fig4} shows the evolution of comoving asymmetries in $\nu_R, \chi_R, \chi_L$ for fixed values of the CP asymmetry parameters, heavy fermion mass and axion decay constant. For this particular benchmark, we choose $\epsilon_{\nu_R}=3 \epsilon, \epsilon_{\chi_R} = -2\epsilon, \epsilon_{\chi_L}=-\epsilon$ with $\epsilon=2.256 \times 10^{-7}$. While the net CP asymmetry is zero by unitarity, the asymmetry generated in $\nu_R$ is consistent with the required lepton asymmetry to generate the observed BAU. Fig. \ref{fig5} shows the parameter space of Dirac ALPy cogenesis in $f-M_1$ plane. The solid contours correspond to the leptogenesis allowed parameter space for two different values of the CP asymmetry parameter. The points represented by the solid blue and maroon colored contours correspond to the parameter space consistent with the observed baryon asymmetry for CP asymmetry parameter values of $\epsilon=2.26 \times 10^{-7}, \epsilon=10^{-2}$ respectively. Similar to the minimal scenario, the shaded region labeled as $\Gamma_{\rm agg}/\mathcal{H}<1$ corresponds to axions not being part of the bath. In this region, calculation of baryon asymmetry requires solving the Boltzmann equation for axions and estimating the non-thermal abundance of heavy neutral fermions, which we do not pursue here. The blue shaded region is ruled out due to strong washouts in equilibrium. The horizontal shaded region at the top leads to axion DM overproduction via vacuum misalignment. Similar to Fig. \ref{fig2b}, here also the dimensionless couplings $\tilde{\lambda}_{i\alpha}, \tilde{h}_{Li}, \tilde{h}_{Ri}$ are assumed to be $\sim 10^{-2}$. The constraints from washouts, $\chi$ relic and $\chi$ lifetime become stronger for larger values of these dimensionless couplings, shrinking the allowed white colored region in Fig. \ref{fig5}.

Depending upon the axion parameter space, $\chi$ can constitute the observed DM entirely or partially. The net dark asymmetry adds up to generate an asymmetric $\chi$ abundance at present epoch $(T=T_0)$ given by 
\begin{equation}
    \Omega_{\chi} (T_0) = m_\chi \frac{(\eta_{\Delta \chi_L}+\eta_{\Delta \chi_R})}{S} \frac{n_\gamma (T_0)}{\rho_{\rm cr}(T_0)}
\end{equation}
where $S$ is the same dilution factor used in Eq. \eqref{eqn:sphconv} and $\rho_{\rm cr}(T_0)=\frac{3\mathcal{H}^2_0}{8\pi G}$ is the critical density of the Universe at present. Since $ \lvert \eta_{\Delta \chi_L}+\eta_{\Delta \chi_R} \rvert \sim \eta_{\Delta \nu_R}$ and $\Omega_{\rm DM} \approx 5.36 \Omega_B$ \cite{Planck:2018vyg}, we get $m_\chi \leq 1.9 m_p$ such that $\Omega_\chi \leq \Omega_{\rm DM}$. However, $\chi$ is not perfectly stable and can have three-body decay with decay width given by 
\begin{equation}
    \Gamma (\chi \rightarrow \nu_R a a) \approx \frac{\tilde{\lambda}^2 \tilde{h}^2}{10240 \pi^3} \frac{m^7_\chi}{f^4 M^2_1}.
\end{equation}
In order to constrain the parameter space in a conservative way, here we consider $m_\chi$ to be same as the one-loop mass given by
\begin{equation}
    m^{\rm loop}_\chi \approx \frac{1}{16\pi^2} \frac{1}{f^2} \tilde{h}_{Li} M^3_{i} \tilde{h}_{Ri}. 
\end{equation}
The green shaded region of Fig. \ref{fig5} shows the region of parameter space in $f-m_\chi$ plane where $\chi$ lifetime $\tau_\chi$ is less than the age of the Universe. Clearly, for realistic values of $f$, particularly for the QCD axion, lifetime criteria for $\chi$ rules out the high mass regime of $N_1$, shown by the green shaded part. On the other hand, the red shaded region of Fig. \ref{fig5} corresponds to the overproduced relic of $\chi$. This leaves only the white region of Fig. \ref{fig5} allowed from successful Dirac ALPy cogenesis and dark matter abundance. It should be noted that the model also allows one-loop $\chi_L-\nu_R$ mixing $m^{\rm loop}_{\chi \nu}$. However, in the limit of sub-eV Dirac neutrino mass $M_D= \frac{y v_2}{\sqrt{2}} \ll m_\chi, m^{\rm loop}_{\chi \nu}$, it does not lead to additional constraints from neutrino data or $\chi$ lifetime requirements.

\begin{figure*}
    \includegraphics[scale=0.25]{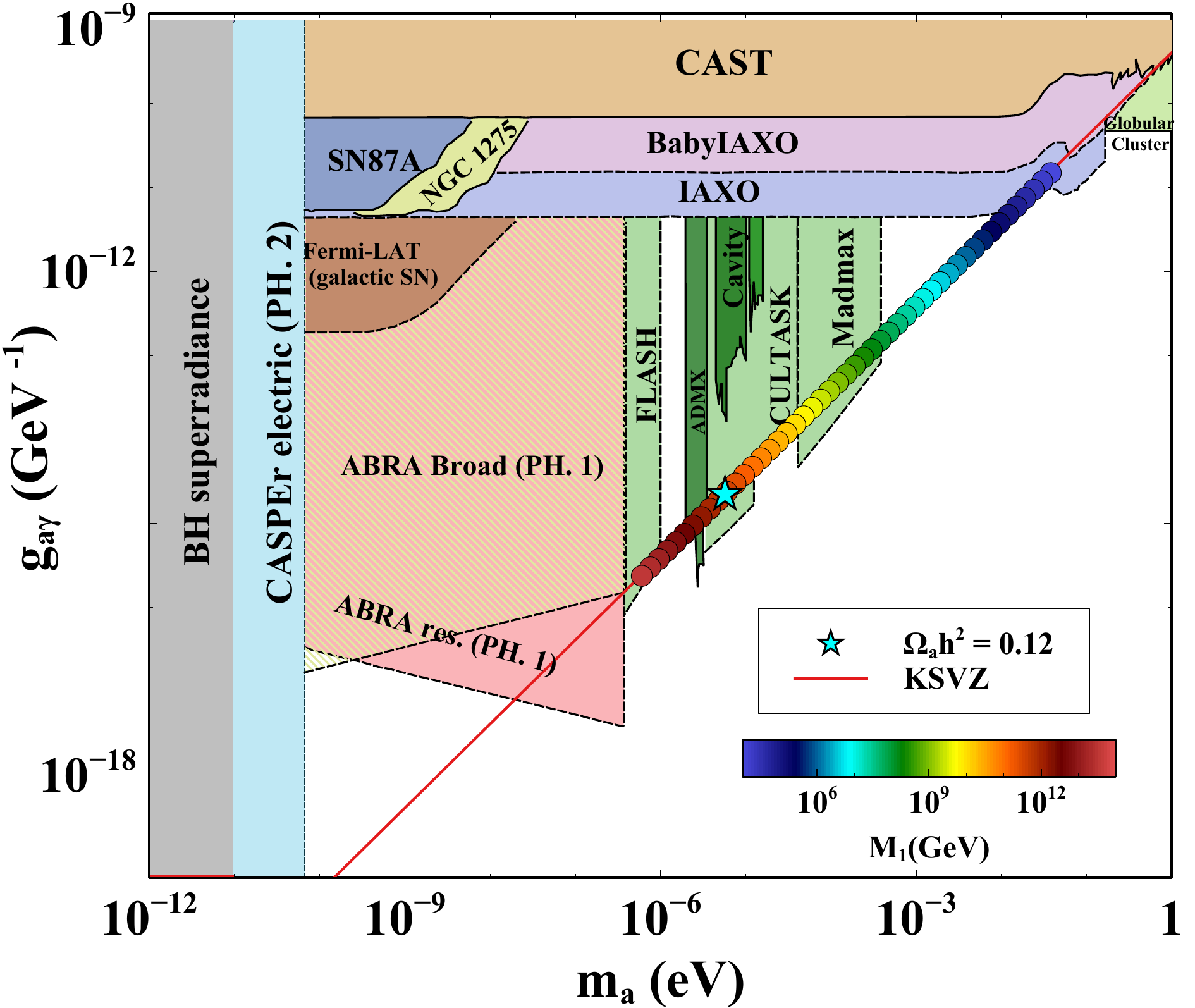}
        \includegraphics[scale=0.25]{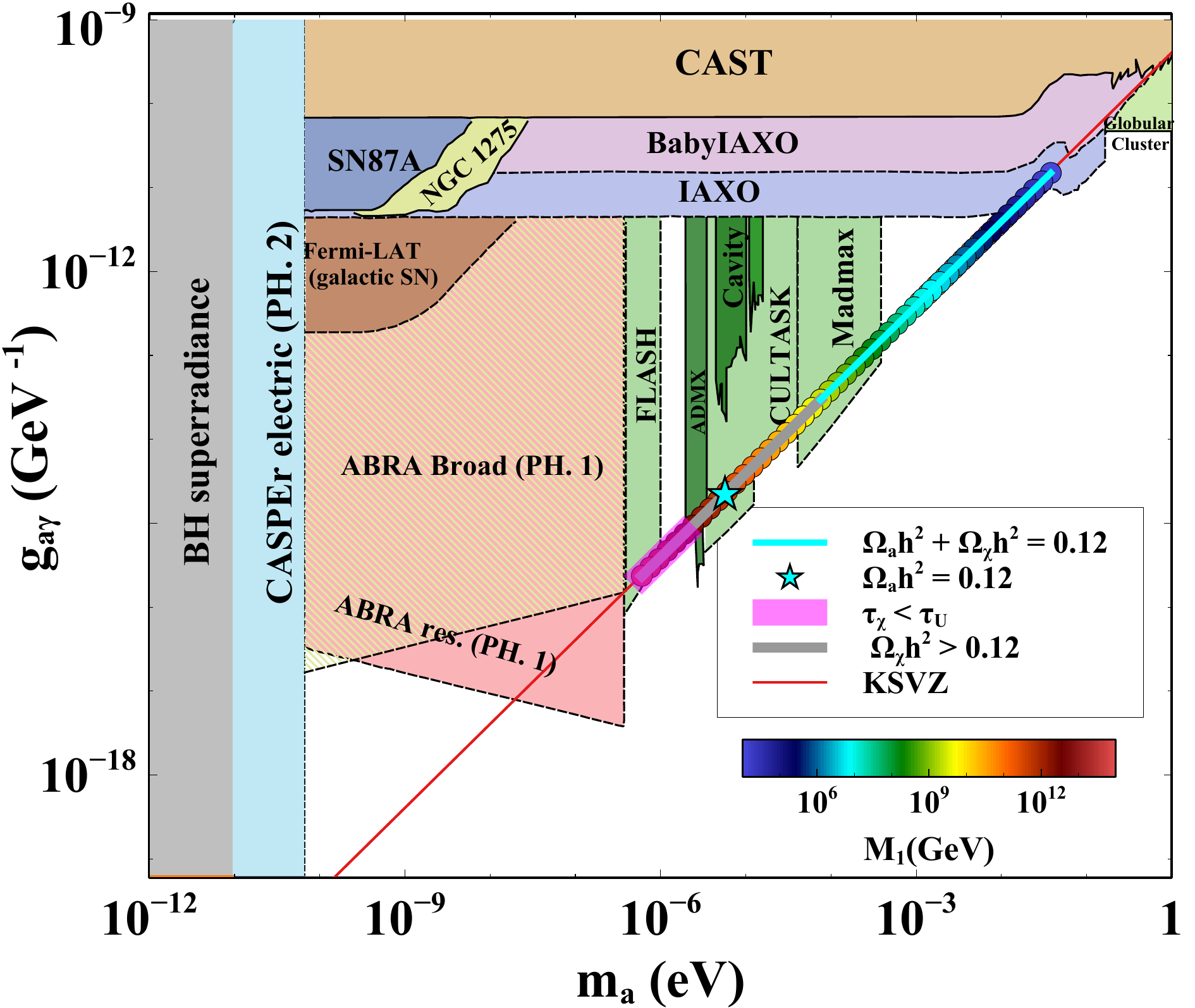}
    \caption{Axion-photon coupling versus axion mass satisfying correct baryon asymmetry for minimal (left) and Dirac (right) ALPy cogenesis. The colored circular points correspond to leptogenesis allowed parameter space for $\epsilon_1=3.67 \times 10^{-7}$ (minimal) and $\epsilon=2.26 \times 10^{-7}$ (Dirac) with color bar indicating the scale of leptogenesis $M_1$. The other parameters are kept same as in Fig. \ref{fig2b} and Fig. \ref{fig5}. The shaded regions correspond to different bounds or sensitivities. The cyan colored $\star$ corresponds to the point where QCD axion DM produced via vacuum misalignment satisfies observed relic while the cyan colored solid line (right panel) corresponds to the region where axion and $\chi$ together generate the observed DM relic. The pink and gray colored bands on the right panel correspond to short-lived $\chi$ compared to the lifetime of the Universe $\tau_U$ and overproduced $\chi$, respectively.}
    \label{fig6}
\end{figure*}

\section{Detection Aspects}
\label{sec3}
As discussed above, leptogenesis in our setup depends upon the heavy fermion mass $M_1$ as well as the axion decay constant $f$. This opens up the possibility to probe ALPy cogenesis favored parameter space indirectly at axion detection experiments. Fig. \ref{fig6} shows the parameter space in the plane of axion-photon coupling $g_{a\gamma}$ and axion mass $m_a$. The current experimental bounds on the axion-photon coupling from various experiments or observables are shown by the solid color lines (CAST \cite{CAST:2007jps, CAST:2017uph}, SN87A \cite{PhysRevLett.60.1797, PhysRevD.39.1020, PhysRevLett.60.1793}, NGC 1275 \cite{Fermi-LAT:2016nkz}, ADMX \cite{ADMX:2006kgb, Stern:2016bbw, ADMX:2019uok}, Globular clusters \cite{Ayala:2014pea}) whereas future experimental sensitivities or observables are shown by the dashed lines (CASPEr \cite{Budker:2013hfa}, FLASH \cite{Alesini:2017ifp,Alesini:2019nzq, Alesini:2023qed}, ABRACADABRA \cite{Kahn:2016aff, Ouellet:2018beu}, CULTASK \cite{Lee:2020cfj, Semertzidis:2019gkj}, MADMAX \cite{Caldwell:2016dcw}, IAXO \cite{Vogel:2013bta, IAXO:2019mpb},  Fermi-LAT \cite{Meyer:2016wrm}, BH superradiance \cite{Cardoso:2018tly}). While for ALP, the axion-photon coupling is not related to axion mass, it is not so for QCD axion. The solid red line in Fig. \ref{fig6} shows the $g_{a\gamma}-m_a$ relation for QCD axion considering the PQ model to be of Kim-Shifman-Vainshtein-Zakharov (KSVZ) \cite{Kim:1979if, Shifman:1979if} type. The circular points on this line correspond to leptogenesis allowed parameter space with the color bar indicating the scale $M_1$. Assuming axion DM to be produced from vacuum misalignment, the minimal ALPy cogenesis is consistent with the observed BAU and DM for the point denoted as cyan colored $\star$ in the left panel plot of Fig. \ref{fig6}. The parameter space widens up to the cyan colored line in the right panel plot of Fig. \ref{fig6} in Dirac ALPy cogenesis as the deficit in axion DM abundance can be filled by asymmetric dark fermion $\chi$. This keeps the parameter space consistent with Dirac ALPy cogenesis verifiable across a wider range of experiments compared to the minimal scenario. In both the plots shown in Fig. \ref{fig6}, we have chosen the values of the CP asymmetry parameter in such a way that a wide range of $f$ remains allowed leading to a wider range of $g_{a\gamma}-m_a$ parameter space within reach of different experiments. For larger values of the CP asymmetry parameter say, $10^{-2}$, $f$ is restricted to small values of $f \lesssim 10^6$ GeV (as seen from Fig. \ref{fig2b} and Fig. \ref{fig5})  which are already ruled out for the QCD axions. For intermediate values of the CP asymmetry parameter, the allowed colored band of circular points in Fig. \ref{fig6} will get squeezed to higher values of $g_{a\gamma}, m_a$.

There exist other detection prospects of our scenario in addition to the ones outlined in Fig. \ref{fig6}. For the choice of our parameters, axions are also produced thermally in the early Universe leaving a hot axion component in addition to the cold axion DM part. For sub-eV axion mass, this leads to component of dark radiation which can be probed via future CMB measurements of effective relativistic degrees of freedom $N_{\rm eff}$. The current bound on such excess dark radiation or $\Delta N_{\rm eff}$ from Planck 2018 measurement reads $\Delta N_{\rm eff} < 0.285$ at $2\sigma$ \cite{Planck:2018vyg}. Future CMB experiments like CMB-S4 and CMB-HD are sensitive to $\Delta N_{\rm eff} = 0.06$ \cite{Abazajian:2019eic} and $\Delta N_{\rm eff} = 0.014$ \cite{CMB-HD:2022bsz} respectively. A thermalized particle like axion with one degree of freedom which decouples from the bath above electroweak scale temperature leads to $\Delta N_{\rm eff} =0.027$ \cite{Borah:2024kfn}. In the Dirac ALPy cogenesis scenario, we also have additional contribution to dark radiation from light Dirac neutrinos. Thermalized light Dirac neutrinos with decoupling temperature above electroweak scale leads to $\Delta N_{\rm eff} =0.14$ \cite{Abazajian:2019oqj}. Therefore, minimal ALPy cogenesis can be probed at future CMB experiments like CMB-HD \cite{CMB-HD:2022bsz} while Dirac ALPy cogenesis remains within reach of multiple CMB missions due to enhanced $\Delta N_{\rm eff}$.

While we have shown the detection prospects of our leptogenesis scenario at axion search and CMB experiments, there are other possible avenues as well. We briefly comment on them without going into the details. Firstly, the minimal ALPy cogenesis scenario has additional heavy neutral leptons (HNL) in sub-TeV to sub-GeV scales due to the Majorana mass of $\nu_R$. Since they mix with active neutrinos, they can have interesting experimental signatures, a recent review of which can be found in \cite{Abdullahi:2022jlv}. HNL below electroweak scale can lead to a variety of final states depending upon their masses and mixing with active neutrinos. There are dedicated ongoing experiments like NA62 \cite{NA62:2017qcd}, FASER \cite{FASER:2018eoc} and upcoming experiments like DUNE \cite{DUNE:2020fgq}, Codex-b \cite{CODEX-b:2019jve}, MATHUSLA\cite{Curtin:2018mvb} to observe such signatures of such light HNL. Additionally, a keV scale HNL can also be a good DM candidate \cite{Drewes:2016upu}. If the scale of such sterile neutrino is even lower, they can also have other interesting implications in neutrino oscillations, neutrinoless double beta decay, astrophysics etc. as summarized in \cite{deGouvea:2006gz}. Secondly, thermal axions at the eV scale also constitute hot dark matter having other indirect detection prospects, as pointed out recently in \cite{Dror:2024ibf}. On the other hand, the long-lived nature of asymmetric DM $\chi$ in Dirac ALPy cogenesis setup can also be probed at indirect search experiments looking for excess photons. Such excess photons can arise via $\chi \rightarrow a a \nu_R$ followed by axion decay or conversion into photons. Finally, both the models discussed here rely on an additional neutrinophilic Higgs doublet which can have a rich phenomenology \cite{Maitra:2014qea, Haba:2011nb, Gabriel:2008es, Seto:2015rma, Bertuzzo:2015ada, Huitu:2017vye}.

\section{Conclusion}
\label{sec4}
We have proposed a novel leptogenesis scenario by considering axion or ALP type interactions involving heavy and light chiral fermions such that the heavy fermion decay into the lighter counterpart and axion can generate a non-zero CP asymmetry. In the first realization of this proposal, we consider two types of chiral fermions $N_R, \nu_R$ such that out-of-equilibrium CP violating decay $N_R \rightarrow \nu_R a$ generates a net asymmetry in $\nu_R$ which later gets transferred to the lepton doublets via Yukawa interactions mediated by a neutrinophilic Higgs doublet. Lepton number violation due to heavy $N_R$ eventually generates Majorana mass of $\nu_R$ at one loop level which results in type-I seesaw origin of light neutrino mass. Demanding this low scale type-I seesaw not to wash out $N_R$-generated asymmetry or generate a new asymmetry, we find the parameter space in terms of the lightest $N_R$'s mass $M_1$ and the axion decay constant $f$. Given astrophysical bounds on $f \gtrsim 10^8$ GeV, we show that successful ALPy cogenesis can occur for $M_1 \gtrsim 10$ TeV. For the QCD axion, this keeps the detection prospects alive at different experiments searching for axions in $\mu$eV-eV range. Interestingly this range also accommodates axion DM. We also propose a Dirac version of the minimal ALPy cogenesis scenario which does not require any net lepton number violation. This not only requires the decaying heavy fermion to be of Dirac type but also require another light Dirac fermion $\chi$. Since asymmetric Dirac fermion $\chi$ can also contribute partially to dark matter in the Universe, the Dirac ALPy cogenesis has a wider parameter space to be probed by axion detection experiments. 

Both the realizations proposed here can be probed at experiments operating at different frontiers. In addition to axion search experiments mentioned above, thermalized axion or light Dirac neutrinos can be probed via precision CMB measurements. Long-lived asymmetric dark matter, new sterile fermions as well as the neutrinophilic Higgs doublet can have rich phenomenology and detection prospects. While the phenomenology of flavor violating ALP couplings to the SM fermions have been studied in the literature \cite{MartinCamalich:2020dfe}, the couplings introduced in ALPy cogenesis here can have additional UV completions and related phenomenology. We leave these details for future studies.

\acknowledgements
The work of D.B. is supported by the Science and Engineering Research Board (SERB), Government of India grants MTR/2022/000575 and CRG/2022/000603. D.B. also acknowledges the support from the Fulbright-Nehru Academic and Professional Excellence Award 2024-25.


\providecommand{\href}[2]{#2}\begingroup\raggedright\endgroup

\end{document}